\pdfoutput=1
\documentclass[letterpaper,11pt,twocolumn]{scrartcl}

\makeatletter
\DeclareOldFontCommand{\rm}{\normalfont\rmfamily}{\mathrm}
\DeclareOldFontCommand{\sf}{\normalfont\sffamily}{\mathsf}
\DeclareOldFontCommand{\tt}{\normalfont\ttfamily}{\mathtt}
\DeclareOldFontCommand{\bf}{\normalfont\bfseries}{\mathbf}
\DeclareOldFontCommand{\it}{\normalfont\itshape}{\mathit}
\DeclareOldFontCommand{\sl}{\normalfont\slshape}{\@nomath\sl}
\DeclareOldFontCommand{\sc}{\normalfont\scshape}{\@nomath\sc}
\makeatother

\usepackage{amsmath,amssymb,color,xcolor,slashed,lineno}
\usepackage[hidelinks]{hyperref}

\PassOptionsToPackage{svgnames}{xcolor}
\usepackage[object=vectorian]{pgfornament} 

\usepackage[top=2.4cm, bottom=2.4cm, left=1.5cm, right=1.5cm,footskip=0.8cm]{geometry}

\usepackage[T1]{fontenc}
\usepackage[utf8]{inputenc}
\usepackage{abstract}

\usepackage[parfill]{parskip}
\usepackage{microtype}
\usepackage[separate-uncertainty = true,multi-part-units=single]{siunitx}
\usepackage{cleveref}
\usepackage{csquotes}
\usepackage{graphicx}
\usepackage{booktabs}
\usepackage{caption}
\usepackage{subcaption}
\usepackage{cuted}
\usepackage{todonotes}

\usepackage[numbers,sort&compress]{natbib}

\usepackage{fancyhdr}
\usepackage[yyyymmdd,hhmmss]{datetime}
\fancypagestyle{plain}
{
}
\usepackage{cleveref}

\usepackage[auth-sc,affil-it]{authblk}

\usepackage{scalefnt}
\newcommand{\abbrev}{\scalefont{.9}}

\newcommand{\MCFM}{\text{\abbrev MCFM}}
\newcommand{\SM}{\text{\abbrev SM}}

\newcommand{\CuTeMCFM}{\text{\abbrev CuTe-MCFM}}
\newcommand{\SCET}{\text{\abbrev SCET}}
\newcommand{\MATRIX}{\text{\abbrev MATRIX}}

\newcommand{\NTHREENNLO}{\text{\abbrev N$^3$LL+NNLO}}
\newcommand{\NTHREENNLOp}{\text{\abbrev N$^3$LL$_\text{p}$+NNLO}}
\newcommand{\NNLL}{\text{\abbrev N$^2$LL}}
\newcommand{\NNNLL}{\text{\abbrev N$^3$LL}}
\newcommand{\NNNLLp}{\text{\abbrev N$^3$LL$_\text{p}$}}
\newcommand{\NNNNLL}{\text{\abbrev N$^4$LL}}
\newcommand{\NNLO}{\text{\abbrev NNLO}}

\newcommand{\NLO}{\text{\abbrev NLO}}
\newcommand{\LHC}{\text{\abbrev LHC}}
\newcommand{\CMS}{\text{\abbrev CMS}}
\newcommand{\ATLAS}{\text{\abbrev ATLAS}}
\newcommand{\LO}{\text{\abbrev LO}}
\newcommand{\PDF}{\text{\abbrev PDF}}
\renewcommand{\d}{\mathrm{d}}
\renewcommand{\tt}{\texttt}
\newcommand{\as}{\ensuremath{\alpha_s}}

\newcounter{notecount}

\makeatletter
\renewcommand\maketitle{
	\begin{center}
		{\huge\bfseries\@title\par\vspace{0.3em}}
		{\scshape\@author, \@date}
	\end{center}
}
\makeatother

\fancypagestyle{firstpage}{%
	\lhead{}
	\rhead{FERMILAB-PUB-22-762-T,IPPP/22/72}
}

\newcommand{\sectionlinetwo}[2]{%
	\nointerlineskip \vspace{.5\baselineskip}\hspace{\fill}
	{\resizebox{0.5\linewidth}{1.2ex}
		{\pgfornament[color = #1]{#2}
	}}%
	\hspace{\fill}
	\par\nointerlineskip \vspace{.5\baselineskip}
}

\begin{document} 

\thispagestyle{firstpage}
\title{\Large{Transverse momentum resummation at \NNNLL+\NNLO{} for diboson processes}}

\author[1]{John M. Campbell}
\author[2]{R. Keith Ellis}
\author[3]{Tobias Neumann}
\author[4]{Satyajit Seth}

\affil[1]{Fermilab, PO Box 500, Batavia IL 60510-5011, USA}
\affil[2]{Institute for Particle Physics Phenomenology, Durham University, Durham, DH1 3LE, UK}
\affil[3]{Department of Physics, Brookhaven National Laboratory, Upton, New York 11973, USA}
\affil[4]{Physical Research Laboratory, Navrangpura, Ahmedabad - 380009, India}

\date{}
\twocolumn[
\maketitle

\vspace{0.5cm}

\begin{onecolabstract}
	\vspace{0.5cm} 
Diboson processes are one of the most accessible and stringent probes of the Standard
Model's electroweak gauge structure at the \LHC{}. They will be probed at the percent level at the 
high-luminosity \LHC{}, challenging current theory predictions. We present transverse 
momentum resummed calculations at \NTHREENNLO{} for the 
processes $ZZ$, $WZ$, $WH$ and $ZH$, compare our predictions with most recent \LHC{} data 
and present predictions at \SI{13.6}{\TeV} including 
theory uncertainty estimates. For $W^+W^-$ production we further present jet-veto resummed results at
\NTHREENNLOp{}.
Our calculations will be made publicly available in the upcoming \MCFM{} 
release and allow future analyses to take advantage of improved predictions.
\end{onecolabstract}
\tableofcontents
\vspace{2em}
]

\section{Introduction}
Large experimental efforts at the \LHC{} are dedicated to the analysis of Standard Model (\SM{})
electroweak 
gauge bosons. The production of $\gamma,W,Z$ and $H$ are typically considered either alone or in 
pairs, see \cref{exptcites} for analyses of diboson processes at \SI{13}{\TeV}. Recent 
developments
include evidence for the triboson processes \cite{CMS:2020hjs,CMS:2019mpq,ATLAS:2019dny}.
The standard treatment of all these processes exploits the
collinear factorization theorem to combine parton distribution functions (\PDF{}s)
and a hard scattering cross-section evaluated at a scale close to $Q$ to derive a prediction.
The scale $Q$ is the invariant mass of the produced colorless final state.
These collinear factorization predictions are not appropriate at small transverse momentum $q_T$,
where predictions at a fixed order of $\as$ contain powers of $L=\log(Q^2/q_T^2)$.
In addition, for the same reason, collinear factorization predictions are not suitable for 
cross-sections where jet
activity is vetoed.
In the region of small transverse momentum the fixed-order predictions need to
be enhanced with resummation of these logarithms to all orders in $\as$. This necessitates
an improved power counting where $\log(Q^2/q_T^2)\sim 1/\as$ and  
exploits a factorization theorem at small $q_T$, valid up to terms suppressed by some power of 
$q_T/Q$.
 
Since the dominant fraction of cross-section resides
at low transverse momentum, accurate theoretical control of this region is important.
In addition, precise resummed predictions are necessary to validate the transverse-momentum spectra 
obtained from parton shower event generators operating at a lower logarithmic accuracy.
Compared to single boson production, resummation effects for boson pair processes are expected to 
be even more important at the same value of $q_T$ because the value of $Q$ is much larger.

Of all massive diboson processes, the production of $W^+W^-$ has received most theoretical and 
phenomenological 
attention. This is because of its sizable cross-section and its role as a background 
to top-quark production and to Higgs-boson production.
Transverse momentum resummation in $W^+W^-$ processes has been
considered in refs.~\cite{Grazzini:2005vw,Meade:2014fca,Bizon:2017rah,Kallweit:2020gva}. In particular
ref.~\cite{Kallweit:2020gva} discusses the
resummation of transverse momentum logarithms at \NNLL{}+\NNLO{}. The
important topic of the resummation of jet veto logarithms in $W^+W^-$
processes has been considered in refs.~\cite{Jaiswal:2014yba,Dawson:2016ysj,Arpino:2019fmo,Kallweit:2020gva}.

As for the other processes, ref.~\cite{Wang:2013qua} considers the $W^\pm Z$ and $ZZ$ processes (as 
well as  $W^+W^-$) at \NNLL+\NLO. Resummation in the $ZZ$ (and $W^+W^-$) processes has been
considered in ref.~\cite{Grazzini:2015wpa} at \NNLL+\NNLO. The interface of 
{\abbrev RadISH} resummation to the \MATRIX{} program allows for \NNNLL{}+\NNLO{} resummation 
\cite{Kallweit:2020gva} of all diboson processes but no phenomenological results for the
$W^\pm Z$ and $ZZ$ processes at this level have been published.

In this paper we present an upgrade of \CuTeMCFM{} \cite{Becher:2020ugp} which implements the 
\SCET{}-based $q_T$ resummation formalism of 
refs.~\cite{Becher:2010tm,Becher:2011xn,Becher:2012yn,Becher:2019bnm}. We describe the \NNNLL{} 
resummation matching to the remaining diboson processes $WW$, 
$ZZ$ and $WZ$ that have been recently implemented in \MCFM{} at fixed order \NNLO{} 
\cite{Campbell:2022gdq}. Our goal is to show these improvements and the phenomenological
capabilities of our code, especially since the diboson calculations were previously only presented 
at a technical level in \MCFM{} \cite{Campbell:2022gdq}.
We present resummed results for the massive diboson processes $W^+W^-$, $W^\pm Z$, and 
$WH$, $ZH$ at the level of \NTHREENNLO{}, compare with data as far as currently available, and 
provide predictions for the current \LHC{} energy of $\sqrt{s}=\SI{13.6}{\TeV}$. 

In addition to $q_T$ resummation, resummation effects become important when we veto against jet 
activity, for example in 
$W^+ W^-$ production to reduce background from $t\bar{t}$ production. 
Although a discussion of jet-veto results is not the principal aim of our study,  
in view of its experimental importance
we present the results of jet-veto resummation for the case of $W^+ W^-$ production. We leave a 
detailed analysis of jet-veto resummation of this and other processes for a future study.

{\centering\sectionlinetwo{black}{88}}

In this paper we use the \SCET{}-based \enquote{collinear anomaly} $q_T$ resummation formalism 
introduced in
refs.~\cite{Becher:2010tm,Becher:2011xn,Becher:2012yn}.
Formulations of $q_T$ resummation that are fully performed in impact parameter space 
have the drawback that the 
transformation
from the impact parameter space $x_T$ back to $q_T$ involve the running coupling at scale $x_T$.
Therefore, when performing the Fourier transform over all values of the impact parameter,
one is forced to introduce a prescription to avoid the Landau pole in the running coupling.
In the formulation of refs.~\cite{Becher:2010tm,Becher:2011xn,Becher:2012yn}
this issue is avoided, setting the scale directly in $q_T$ space.
The cross-section is obtained by combining the contributions from the partonic channels 
$i,j\in{q,\bar{q},g}$.
Up to terms suppressed by powers of $q_T/Q$, these channels exhibit a factorized form that is fully differential in the
momenta $\{q\}$ of the colorless final state
\begin{table}[t]
	\caption{Experimental publications for boson pair production at \SI{13}{\TeV}.}
	\vspace{0.5em}
	\begin{center}
		\begin{tabular}{|l|l|l|}
			\hline
			Process & ATLAS & CMS \\
			\hline
			$WZ$&\cite{ATLAS:2019bsc}&\cite{CMS:2019efc,CMS:2019ppl,CMS:2021icx}\\
			$ZZ$&\cite{ATLAS:2019xhj,ATLAS:2021kog}&\cite{CMS:2020gtj}\\
			$WW$&\cite{ATLAS:2017bbg,ATLAS:2019rob}&\cite{CMS:2019ppl,CMS:2020mxy}\\
			$WH/ZH$ & \cite{ATLAS:2020fcp,ATLAS:2020jwz}&\cite{CMS:2018nsn}\\
			\hline
		\end{tabular}
	\end{center}
	\label{exptcites} 
\end{table}
\begin{multline}
\d\sigma_{ij}(p_1,p_2,\{\underline{q} \}) = \\
 \int_0^1 \d z_1 \int_0^1 \d z_2\,
\d\sigma^0_{ij}(z_1 p_1,z_2 p_2,\{\underline{q}\})\,
\mathcal H_{ij}(z_1 p_1,z_2  p_2,\{\underline{q} \},\mu) \,\\
\times \frac{1}{4\pi}\int\d^2x_\perp \,
e^{-iq_\perp x_\perp}
\left(\frac{x_T^2 Q^2}{b_0^2}\right)^{-F_{ij}(x_\perp,\mu)} \\
 \times B_i(z_1,x_\perp,\mu) \cdot
B_j(z_2,x_\perp,\mu)\,,
\label{eq:fact}
\end{multline}

where $p_1$ and $p_2$ are the incoming hadron momenta.
The function
$\d\sigma^0_{ij}$ denotes the differential cross-section for
the hard Born-level process and the hard-function $\mathcal{H}_{ij}$ contains the associated
virtual corrections. The beam functions $B_i$ and $B_j$ include the
effects of soft and collinear emissions at large transverse separation $x_\perp$
and the indices $i$ and $j$ and the momentum fractions $z_1$ and
$z_2$ refer to the partons which enter the hard process after these
emissions.  The collinear anomaly leads to the $Q^2$-dependent factor
within the Fourier-integral over the transverse position
$x_\perp$. The perturbatively calculable anomaly exponent $F_{ij}$ is
also referred to as the rapidity anomalous dimension in the framework
of ref.~\cite{Chiu:2012ir}.  We further have $b_0=2e^{-\gamma_E}$,
where $\gamma_E$ is the Euler constant, and
$x_T^2=-x_\perp^2$. 

This framework for $q_T$ resummation has been implemented at \NNNLL{} in \CuTeMCFM{} 
\cite{Becher:2020ugp,Neumann:2021zkb}, see ref.~\cite{Becher:2020ugp} for further details. 
Matching to large-$q_T$ 
fixed-order predictions were previously performed at relative
order $\alpha_s^2$ for the processes $H,Z,W^\pm$~\cite{Boughezal:2016wmq}, $W^\pm H$ and
$ZH$~\cite{Campbell:2016jau}, $\gamma\gamma$~\cite{Campbell:2016yrh},
$Z\gamma$~\cite{Campbell:2017aul}, as well as at {\abbrev N$^4$LL+N$^3$LO} for $Z$ production in 
ref.~\cite{Neumann:2022lft}. The code is fully differential in the Born
kinematics, including the decays of the bosons and provides an
efficient way to estimate uncertainties from fixed-order truncation,
resummation, and parton distribution functions.

To provide phenomenologically meaningful results also for $W^+W^-$ production, we have implemented  
jet-veto resummation at \NTHREENNLOp{} following the collinear anomaly formalism of 
ref.~\cite{Becher:2013xia}. Beam- and soft-functions are taken from 
refs.~\cite{Abreu:2022sdc,Abreu:2022zgo} and the rapidity anomalous dimension at the two-loop level 
is taken from refs.~\cite{Banfi:2012jm,Becher:2013xia}.
The notation \NNNLLp{} indicates that full \NNNLL{} accuracy is not achieved since an approximate
form, valid at small jet-radii, for the three-loop term in the collinear anomaly exponent is used~\cite{Banfi:2015pju}.
A detailed presentation of our implementation and its phenomenology for various processes will be 
presented elsewhere \cite{CENS}.

\section{Phenomenology}
In the following we first present finely binned transverse momentum spectra at $\SI{13.6}{\TeV}$ 
and 
compare fixed-order and resummation improved predictions for each diboson process. These 
demonstrate the 
impact of the \NNNLL{} resummation for future analyses. In practice, in current experimental 
analyses the binning is still large, so that the impact of resummation is less apparent. We compare 
with experimental measurements as far as available for the 13~TeV \LHC{}.

In \cref{sec:ZZproduction} we first consider $ZZ$ production. For this process the transverse 
momentum of the vector boson pair system is directly measured, unlike for processes with $W$ 
bosons which have missing energy. We compare with differential and total cross-section measurements 
from both \CMS{} and 
\ATLAS{}.
We then present results for $W^\pm Z$ production in \cref{sec:WZproduction} and compare with 
\ATLAS{} data. Finally we present jet-veto resummed predictions for $W^+W^-$ in 
\cref{sec:WWproduction} and compare to \CMS{} measurements. Finally, we show  differential 
predictions at \SI{13.6}{\TeV} for $W^\pm H$ and $ZH$ in \cref{sec:VHproduction}.

\paragraph{Input parameters.}
Throughout this paper we use the \PDF{} set
{\tt{ NNPDF31\_nnlo\_as\_0118}}
which has five active flavors, except for the $W^+W^-$ process where we use the \PDF{} set {\tt{ 
		NNPDF31\_nnlo\_as\_0118\_nf\_4}}
with four active flavors \cite{NNPDF:2017mvq}. We work in the electroweak $G_\mu$ scheme with $m_W 
= \SI{80.385}{\GeV}$, $m_Z=\SI{91.1876}{\GeV}$, $G_\mu=\SI{1.166390e-5}{\GeV^{-2}}$ and further 
have $\Gamma_W=\SI{2.0854}{\GeV}$, $\Gamma_Z=\SI{2.4952}{\GeV}$, $m_H=\SI{125}{\GeV}$, $m_t=\SI{173.3}{\GeV}$. 

At fixed 
order we set 
the default renormalization and factorization scales to the invariant mass of the diboson system.
For the resummation-improved results we vary hard scale, resummation scale and rapidity scale 
following refs.~\cite{Becher:2020ugp,Neumann:2021zkb}. We symmetrize resummation uncertainty bands 
to 
account for a frozen out downwards scale variation at small $q_T$ that would otherwise evaluate 
$\alpha_s$ in the non-perturbative regime. Since our resummation includes the matching through a 
transition function, we vary this function to estimate a matching uncertainty and include this in 
the uncertainty bands. For the detailed procedure we refer the reader to ref.~\cite{Becher:2020ugp}.

\subsection{$ZZ$ production} \label{sec:ZZproduction}

\subsubsection{$ZZ$ production at $\sqrt{s}=\SI{13.6}{\TeV}$}

\begin{table}
	\caption{\label{tab:ZZ-CMS} Fiducial volume of the
		\CMS{} $ZZ$ analysis presented in ref.~\cite{CMS:2020gtj}.}
	\vspace{0.5em}
	\begin{center}
		\begin{tabular}{r | l }
			{ lepton cuts} & $q_T^{l_1} > \SI{20}{\GeV}$, $q_T^{l_2} > \SI{10}{\GeV}$, \\
			&  $q_T^{l_{3,4}} > \SI{10}{\GeV}$, $|\eta^{l}|<2.5$ \\
			{lepton pair mass}& $\SI{60}{\GeV}<m_{l^-l^+}<\SI{120}{\GeV}$
		\end{tabular}
	\end{center}
\end{table}

\begin{figure}
	\centering
	\includegraphics[width=0.85\columnwidth]{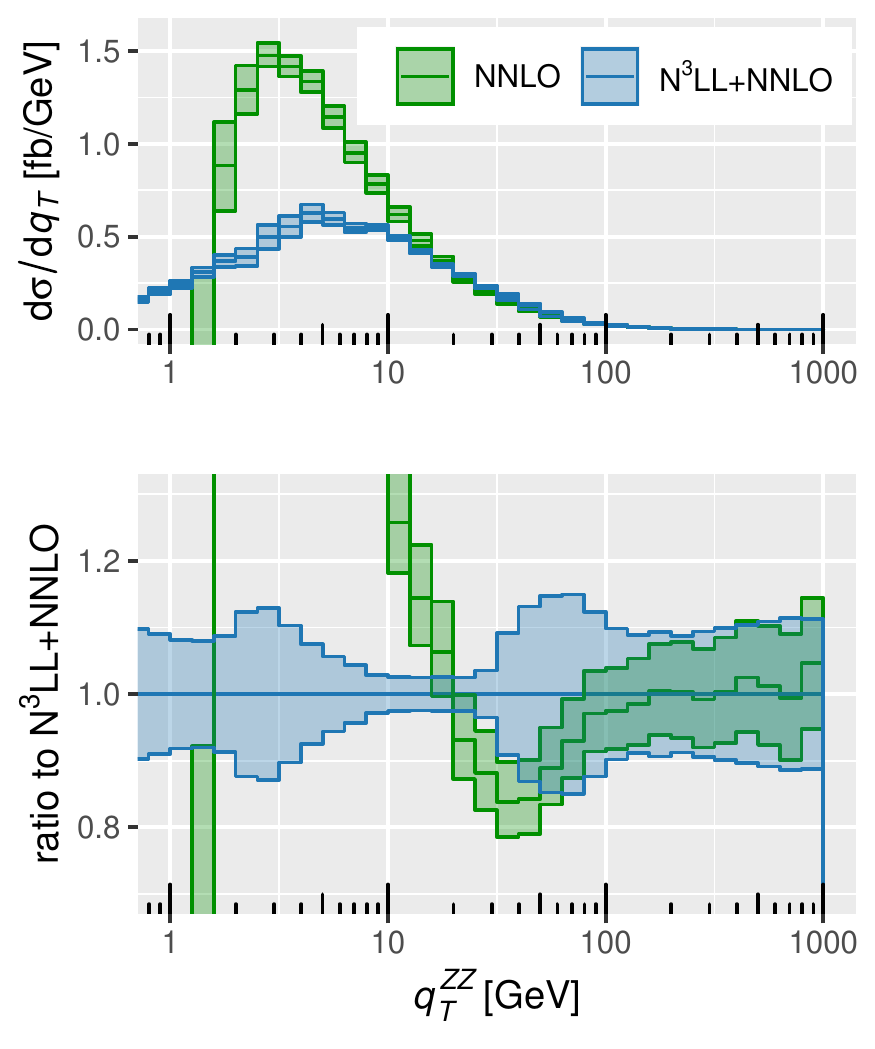}
	\caption{Transverse momentum distribution of the $ZZ$ pair at \NNLO{} and \NNNLL{}+\NNLO{} 
	using the  \CMS{} 
	cuts in \cref{tab:ZZ-CMS}  \cite{CMS:2020gtj}, but at \SI{13.6}{\TeV}. }
	\label{fig:ZZ136qt}
\end{figure}

We first present results for $ZZ$ production at $\sqrt{s}=\SI{13.6}{\TeV}$ using the \CMS{}
cuts in \cref{tab:ZZ-CMS} \cite{CMS:2020gtj} to study the impact of the resummation compared to 
fixed order. In 
\cref{fig:ZZ136qt} we show the $ZZ$ transverse momentum distribution at \NNLO{} fixed-order and 
matched with \NNNLL{} resummation. 
The transition region is around \SIrange{30}{100}{\GeV} and leads to uncertainties of about 15\% 
in that region, comparable to the fixed-order uncertainties of 10\%. The uncertainties in the 
resummation region for smaller $q_T$ benefit from the high logarithmic accuracy until very small 
$q_T$ of about \SIrange{4}{5}{\GeV}. Here resummation at the level of \NNNNLL{} would improve 
uncertainties further \cite{Neumann:2022lft}. Overall we conclude that resummation within current 
theory uncertainty levels becomes important below about \SIrange{50}{100}{\GeV}.

\subsubsection{Comparison with \CMS{} measurements}
We compare our predictions with the $\sqrt{s}=\SI{13}{\TeV}$ \CMS{} results of 
ref.~\cite{CMS:2020gtj}.
The cuts for our analysis are shown in \cref{tab:ZZ-CMS}. To simplify our theoretical analysis we
perform our calculations for $Z$ bosons decaying to different-flavor leptons
and account for all combinations with an overall factor of two.  We have checked that this
results in a negligible difference in our results at \NLO{}. We neglect
identical-particle effects (i.e. the $e^- e^+ e^- e^+$ final state is treated in the same
way as $e^- e^+ \mu^- \mu^+$).

\begin{figure}
	\centering
\includegraphics[width=0.85\columnwidth]{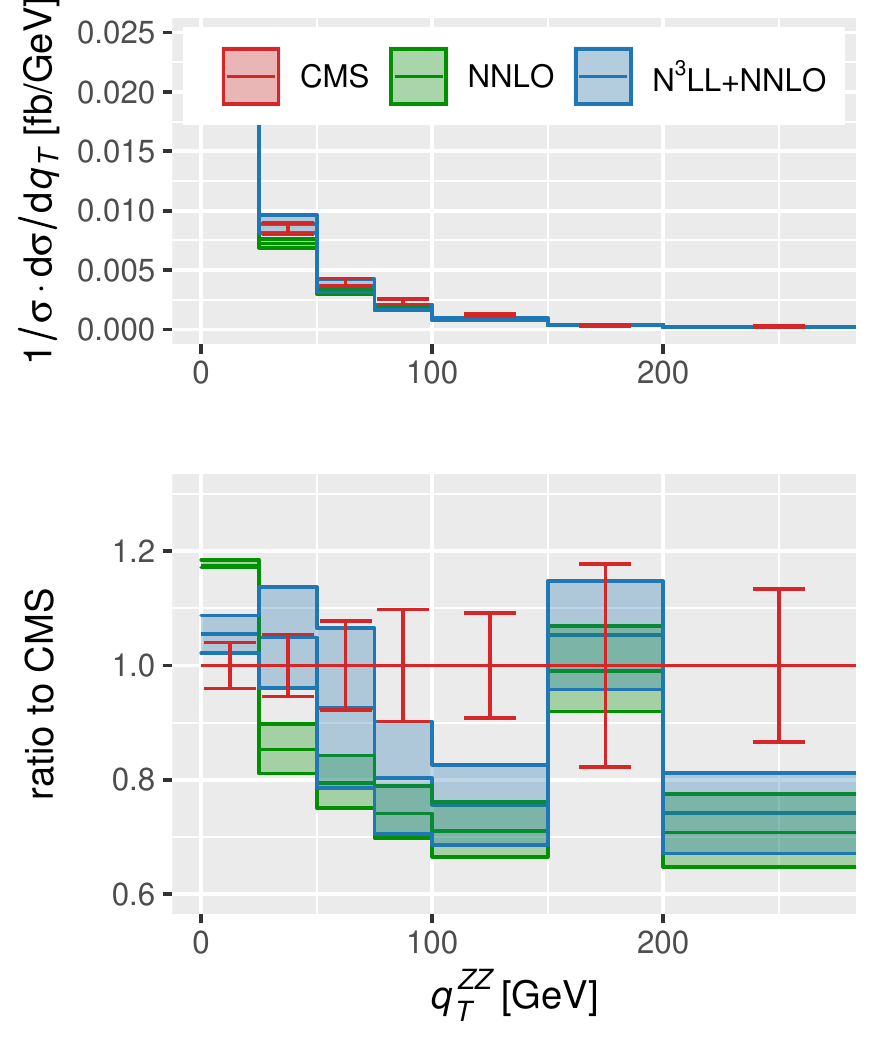}
\caption{
	The $q_T^{ZZ}$ distribution at 
	 \NNLO{} and \NNNLL{}+\NNLO{} compared to the \CMS{} data
from ref.~\cite{CMS:2020gtj}.
}
\label{fig:ZZCMSptZZ}
\end{figure}

A comparison of the fixed-order \NNLO{} and resummed \NTHREENNLO{} predictions for $q_T^{ZZ}$ is 
shown in 
\cref{fig:ZZCMSptZZ}, also compared
with the corresponding \CMS{} data (c.f. Fig.~5 (left) of ref.~\cite{CMS:2020gtj}).
The resummation improves the description of the experimental data up to \SI{75}{\GeV} noticeably, 
as anticipated by our finely binned analysis in 
\cref{fig:ZZ136qt}.

\begin{figure}
	\centering
	\includegraphics[width=0.85\columnwidth]{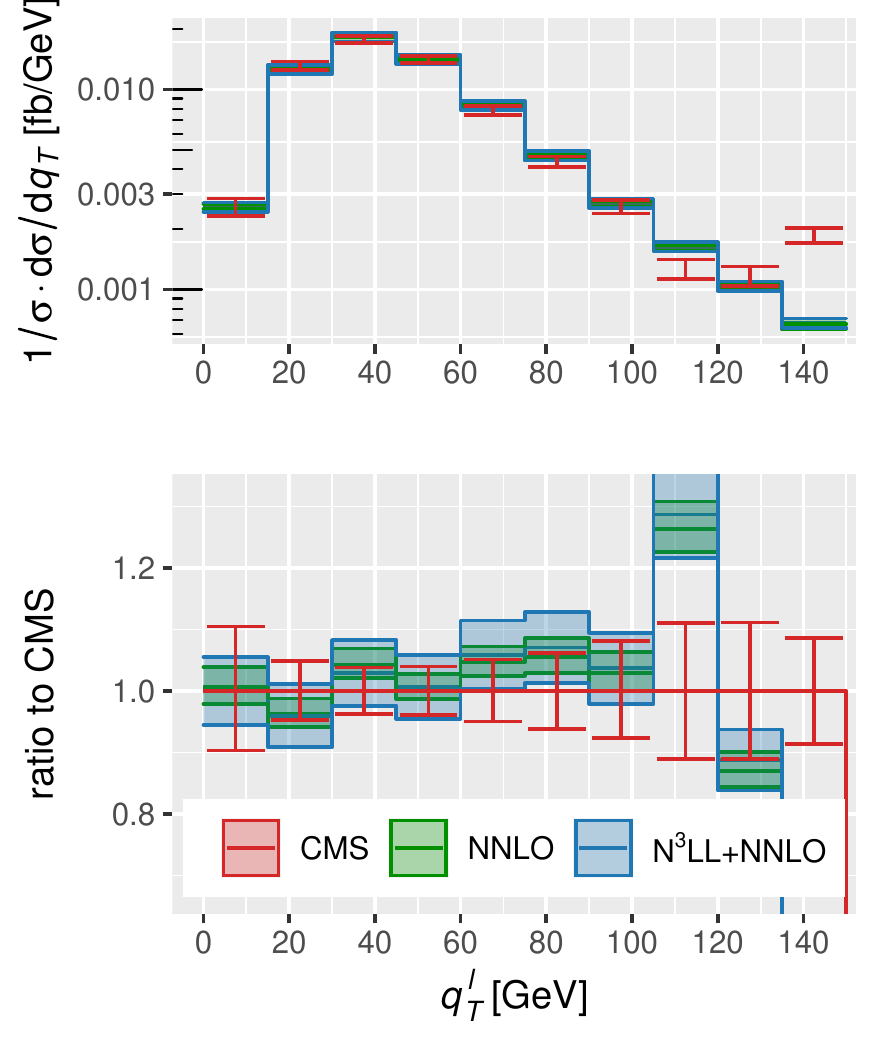}
	\caption{The $q_T^l$ (summed over all leptons) distribution at
		\NNLO{} and \NNNLL{}+\NNLO{}, compared to the \CMS{} data from ref.~\cite{CMS:2020gtj}.}
	\label{fig:ZZCMSptlall}
\end{figure}

\begin{figure}
	\centering
	\includegraphics[width=0.85\columnwidth]{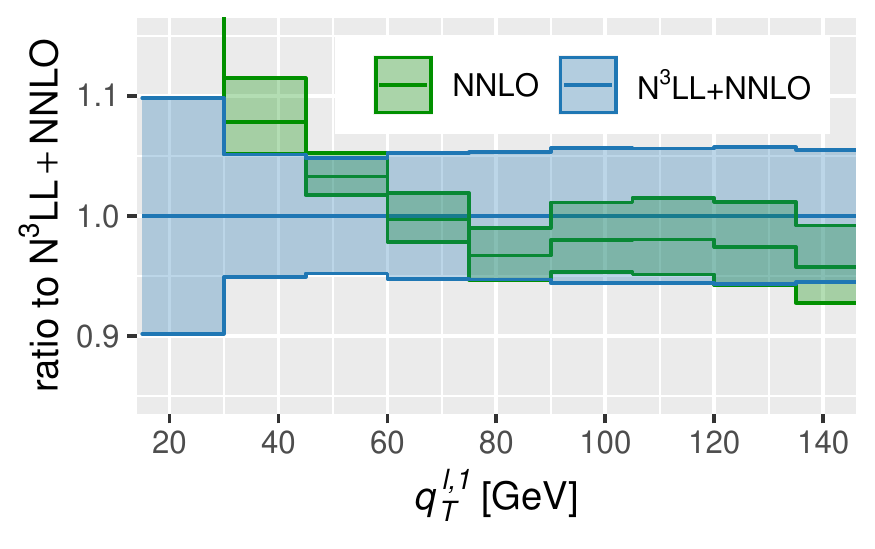}
	\caption{The $q_T^{\ell_1}$ distribution at \NNLO{} as a ratio to \NNNLL{}+\NNLO{}. See 
		\cref{fig:ZZCMSptlall} for the sum of all leptons compared to \CMS{} data.}
	\label{fig:ZZCMSptl1}
\end{figure}

The \CMS{} collaboration also presents a measurement of the transverse momentum of all four 
leptons, see fig.~4 (left) of ref.~\cite{CMS:2020gtj}.  Our
\NNLO{} and \NTHREENNLO{} results for this distribution are shown in \cref{fig:ZZCMSptlall}.  In 
this distribution there is no evidence for the effects of resummation.  Our theoretical
study of the leading-lepton $q_T^{l,1}$ spectrum shown in \cref{fig:ZZCMSptl1}
displays the importance of resummation effects, but only at very small $q_T$ with large 
uncertainties. We have checked
that the distributions of the other leptons are not significantly changed by $q_T$ resummation,
which leads to this effect being washed out in the experimental measurement.

Last, we compare the total 
fiducial cross-section prediction with the measurement in \cref{tab:ZZcrossCMS} and find reasonable 
agreement between theory prediction and measurement.

\begin{table}[]
	\caption{Comparison of total fiducial $ZZ$ cross-section predictions at \NNLO{}, \NTHREENNLO{} 
	with 
		the \CMS{} analysis combining measurements from 2016, 2017 and 2018 \cite{CMS:2020gtj}. 
		Fiducial cuts are as in \cref{tab:ZZ-CMS}. }
	\label{tab:ZZcrossCMS}
	\begin{tabular}{@{}ll@{}}
		\toprule
		& cross-section [fb] \\ \midrule
		\NNLO{}        &      $37.8^{+0.5}_{-0.4} \, \text{(scale)} $                  \\
		\NTHREENNLO{}     &   $36.0 \pm 0.8\,\text{(scale)} \pm 
		0.8\,\text{(match)}$                     \\
		measurement &  $40.5\pm0.7\,\text{(sta.)} \pm 1.1\, \text{(sys.)} \pm 0.7\, 
		\text{(lum.)}$               \\ \bottomrule
	\end{tabular}
\end{table}

\subsubsection{Comparison with \ATLAS{} measurements}

We also compare with results from the ATLAS collaboration~\cite{ATLAS:2021kog} using
the cuts shown in \cref{tab:ZZ-ATLAS}.  As before,  we
perform our calculations for $Z$ bosons decaying to different-flavor leptons
and account for all combinations with an overall factor of two.

\begin{table}
	\caption{\label{tab:ZZ-ATLAS} Setup for the 
		\ATLAS{} $ZZ$ analysis at $\sqrt{s}=\SI{13}{\TeV}$ presented in ref.~\cite{ATLAS:2021kog}.}
	\vspace{0.5em}
	\begin{center}
		\begin{tabular}{r | l }
			{ lepton cuts}
			& $q_T^{\ell_1} > \SI{20}{\GeV}$, $q_T^{\ell_2} > \SI{10}{\GeV}$, \\
			& $q_T^{\ell_{3,4}} > \SI{5}{\GeV}$, $q_T^{e} > \SI{7}{\GeV}$,\\
			& $|\eta^{\mu}|<2.7$, $|\eta^{e}|<2.47$\\
			{lepton separation}& $\Delta R(\ell,\ell^\prime) > 0.05$ 
		\end{tabular}
	\end{center}
\end{table}

\begin{figure}
	\centering
	\includegraphics[width=0.85\columnwidth]{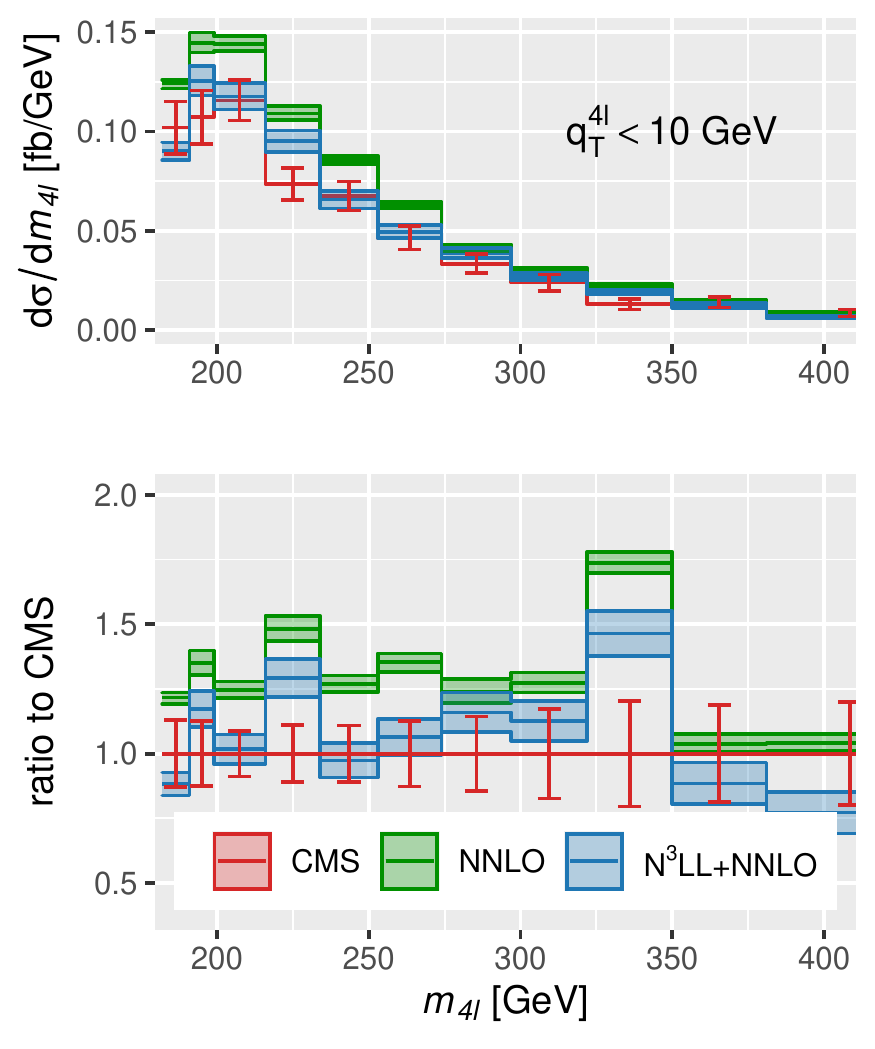}
\caption{The $m_{4l}$ distribution for $q_{T}^{4\ell}<\SI{10}{\GeV}$ at \NNLO{} and \NTHREENNLO{} 
compared with \ATLAS{} data from ref.~\cite{ATLAS:2021kog}. }
\label{fig:ZZATLASNNLO}
\end{figure}

The \ATLAS{} collaboration has performed measurements of the
$m_{4l}$ distribution in five slices of $q_T^{4\ell}$
in fig.~15 of ref.~\cite{ATLAS:2021kog}. 
We limit our comparison to the region $m_{4\ell} > \SI{182}{\GeV}$ to avoid the low invariant mass 
region populated by $gg\to H$.
Since we are resumming logarithms $\log (m_{4\ell}/q_{T}^{4\ell})$ our expectation
is that the resummation should improve the agreement with data in the region of small 
$q_T^{4\ell}$, 
in particular as $m_{4\ell}$ increases. We show results at \NNLO{} and \NTHREENNLO{} in
\cref{fig:ZZATLASNNLO} and indeed find this expectation to be correct. For brevity we only show the 
comparison with the first slice $q_T^{4\ell}<\SI{10}{\GeV}$.

\begin{table}[]
	\caption{Comparison of total fiducial $ZZ$ cross-section predictions in the on-shell region 
		$\SI{180}{\GeV} < m_{4l} < \SI{2000}{\GeV}$ at \NNLO{}, \NTHREENNLO{} with 
		the \ATLAS{} analysis \cite{ATLAS:2021kog}. Fiducial cuts are as in \cref{tab:ZZ-ATLAS}. }
	\label{tab:ZZcrossATLAS}
	\begin{tabular}{@{}ll@{}}
		\toprule
		& cross-section [fb] \\ \midrule
		\NNLO{}        &      $45.3^{+1.1}_{-0.9}$~fb                  \\
		\NTHREENNLO{}     &   $43.7 \pm 0.7\, \text{(scale)} \pm 
		0.8\,\text{(match)}$                     \\
		measurement &  $49.3\pm0.8\,\text{(sta.)} \pm 0.8\, \text{(sys.)} \pm 0.8\, 
		\text{(lum.)}$               \\ \bottomrule
	\end{tabular}
\end{table}

\subsection{$W^\pm Z$ production} \label{sec:WZproduction}

\subsubsection{$WZ$ production at $\sqrt{s}=\SI{13.6}{\TeV}$}

We begin with predictions at $\SI{13.6}{\TeV}$ for run~3 of the \LHC{} using \CMS{} cuts as in
\cref{tab:WZ-CMS}. 

\begin{figure}
	\centering
	\includegraphics[width=0.85\columnwidth]{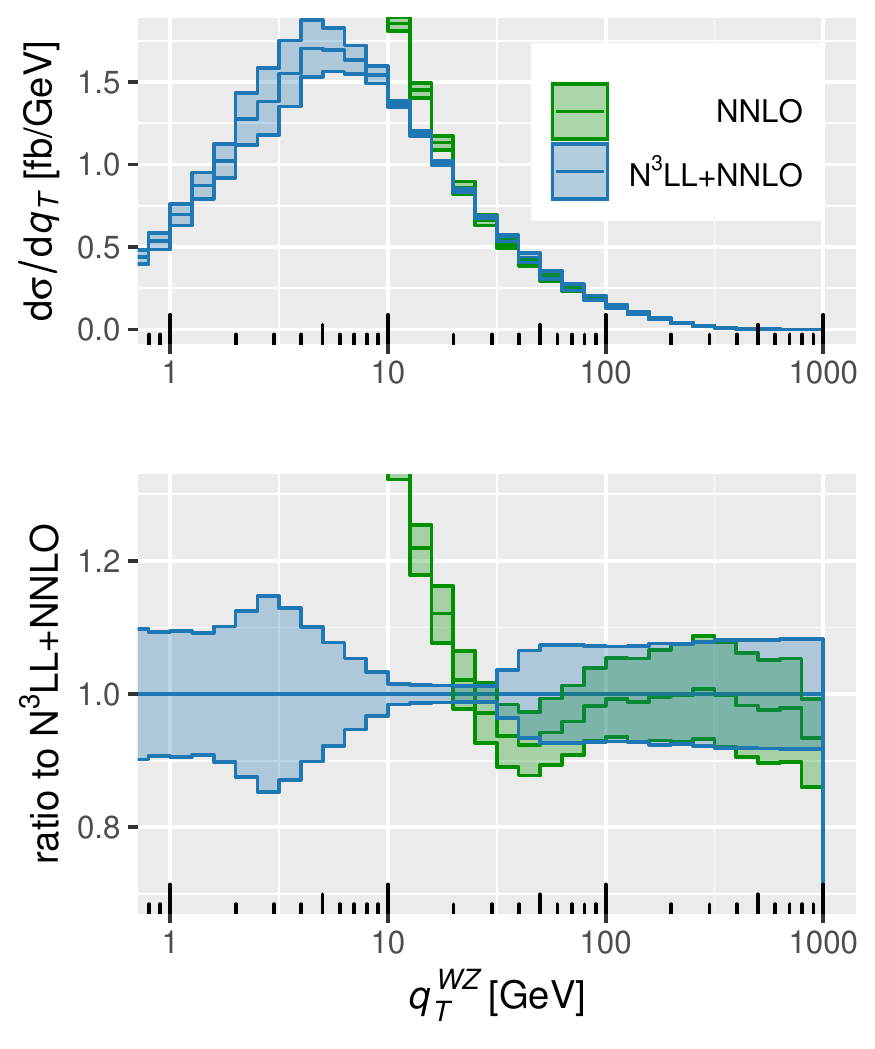}
	\caption{Comparison of \NNLO{} and \NNNLL{}+\NNLO{} predictions for truth $p_T^{W^\pm Z}$ at 
	\SI{13.6}{\TeV} using the \CMS{} cuts in \cref{tab:WZ-CMS}.}
	\label{fig:WZ136qt}
\end{figure}

Fig.~\ref{fig:WZ136qt} illustrates the impact that resummation has on the
$q_T$ distribution.  For the purposes of illustration, $q_T$ is constructed
from the full $WZ$ four-vector, although of course this is not a quantity that
can be directly measured in experiment. Similar to the other diboson processes, the resummation 
becomes essential below \SIrange{50}{100}{\GeV}.

\begin{figure}[t]
	\centering
	\includegraphics[width=0.85\columnwidth]{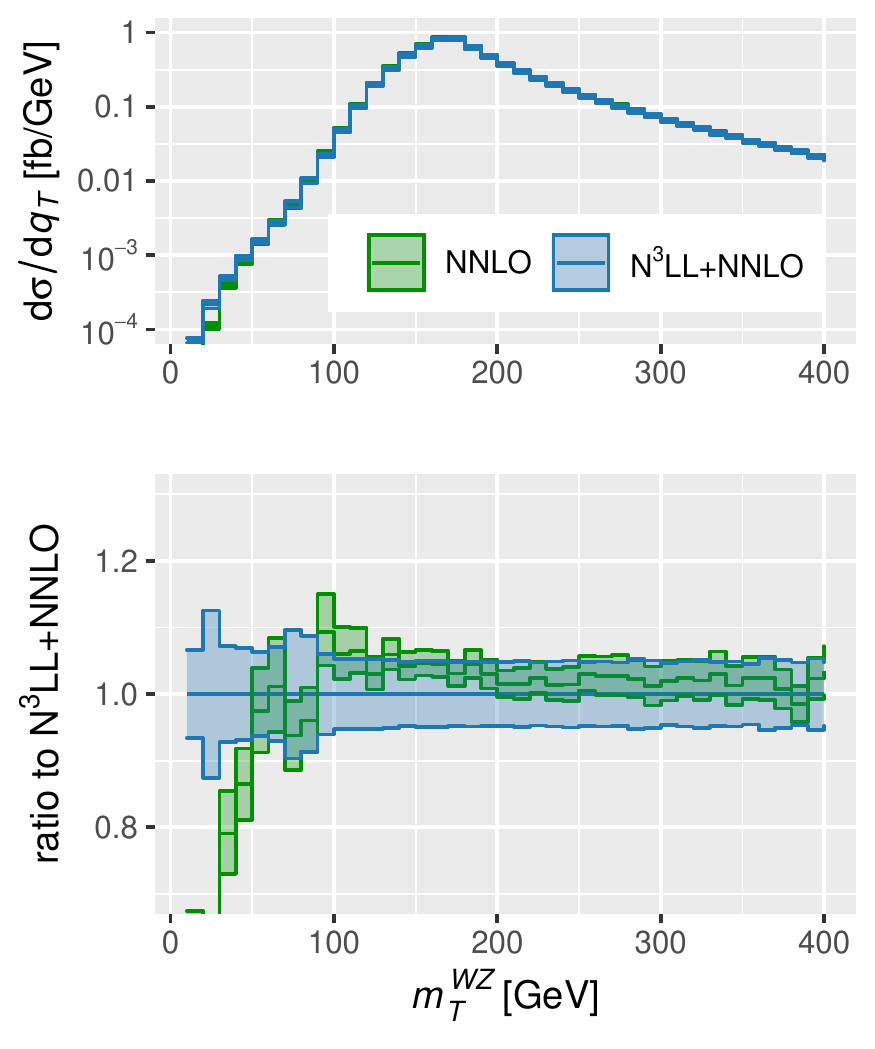}
	\caption{Comparison of \NNLO{} and \NTHREENNLO{} predictions for the truth $m_T^{W^\pm Z}$ at 
	\SI{13.6}{\TeV} using the \CMS{} cuts in \cref{tab:WZ-CMS}.}
	\label{fig:WZ136mT}
\end{figure}

A related quantity, which is often measured in experiment, is the 
transverse mass of the $WZ$ system, $m_T^{WZ}$, which following ref.~\cite{ATLAS:2019bsc} is defined as,
\begin{multline}
	\left(m_{\mathrm{T}}^{WZ}\right)^2 = \left( \sum_{\ell = 1}^3 p_{\mathrm{T}}^\ell +
	E_{\mathrm{T}}^{\mathrm{miss}} \right)^2 \\
	- \left[ \left(\sum_{\ell = 1}^3 p_x^\ell + E_{x}^{\mathrm{miss}} \right)^2 + 
	\left(\sum_{\ell = 1}^3 p_y^\ell + E_{y}^{\mathrm{miss}} \right)^2 \right]  
	\,.
	\label{eq:mT}
\end{multline}
The predictions for this variable are shown in 
\cref{fig:WZ136mT}. At the current level of theory uncertainties, resummation effects are relevant 
for transverse masses less than about \SI{100}{\GeV}, far below the peak region.

\subsubsection{Comparison with \CMS{} measurements}

For $W^\pm Z$ production, we choose to focus on the \CMS{} analysis of  ref.~\cite{CMS:2021icx}.
The parameters and cuts for this study are given in \cref{tab:WZ-CMS}.
As in previous sections, we slightly simplify the theoretical analysis by computing the
cross-section for different-flavor leptons only.
We sum over lepton flavors by applying an overall factor
of four, neglecting same-flavor effects at the $2\%$ level~\cite{CMS:2021icx} that are, however,
unimportant for the current experimental and theoretical accuracy.
We first compare total fiducial cross-sections in \cref{tab:WZcrossCMS} 
and find agreement between theory prediction and measurement within uncertainties.
\begin{table}
	\caption{\label{tab:WZ-CMS} Setup for the 
		\CMS{} $WZ$ fiducial volume analysis at $\sqrt{s}=\SI{13}{\TeV}$ presented in 
		ref.~\cite{CMS:2021icx}.}
	\vspace{0.5em}
\begin{center}
\begin{tabular}{r | l }
{ lepton cuts}
& $p_T^{\ell_{Z1}} > \SI{25}{\GeV}$, $p_T^{\ell_{Z2}} > \SI{10}{\GeV}$, \\
 & $p_T^{\ell_{W}} > \SI{25}{\GeV}$,   $|\eta^{\ell}|<2.5$, \\ 
&  $\SI{60}{\GeV}<m_{\ell^-\ell^+}<\SI{120}{\GeV}$, \\
& $M(\ell_{Z1},\ell_{Z2},\ell_W) > \SI{100}{\GeV}$
\end{tabular}
\end{center}

\end{table}

\begin{table*}
	\centering
	\caption{Comparison of total fiducial $WZ$ cross-section predictions at \NNLO{}, \NTHREENNLO{} 
	with the \CMS{} analysis \cite{CMS:2021icx}. Fiducial cuts are as in \cref{tab:WZ-CMS}. }
	\label{tab:WZcrossCMS}
	\begin{tabular}{@{}lll@{}}
		\toprule
		& $W^-Z \to e^-\bar\nu \mu^-\mu^+$  & $W^+Z \to e^+\nu \mu^-\mu^+ $ \\ \midrule
		\NNLO{}       &      $29.9^{+0.7}_{-0.6}$      & $42.8^{+0.9}_{-0.9}$            \\
		\NTHREENNLO{}   & $29.0\pm1.1\,\text{(scale)}\pm0.3\,\text{(match.)}$  & 
		$41.6\pm1.5\,\text{(scale)}\pm0.4\,\text{(match.)}$              
		\\
		measurement &  
		$31.8\pm1.4\,\text{(sta.)} \pm 1.1\, \text{(sys.)} \pm 0.6\, \text{(lum.)}$  &
		$43.1\pm1.4\,\text{(sta.)} \pm 1.5\, \text{(sys.)} \pm 0.9\, \text{(lum.)}$  \\ \bottomrule
	\end{tabular}
\end{table*}

\begin{figure}[t]
	\centering
	\includegraphics[width=0.85\columnwidth]{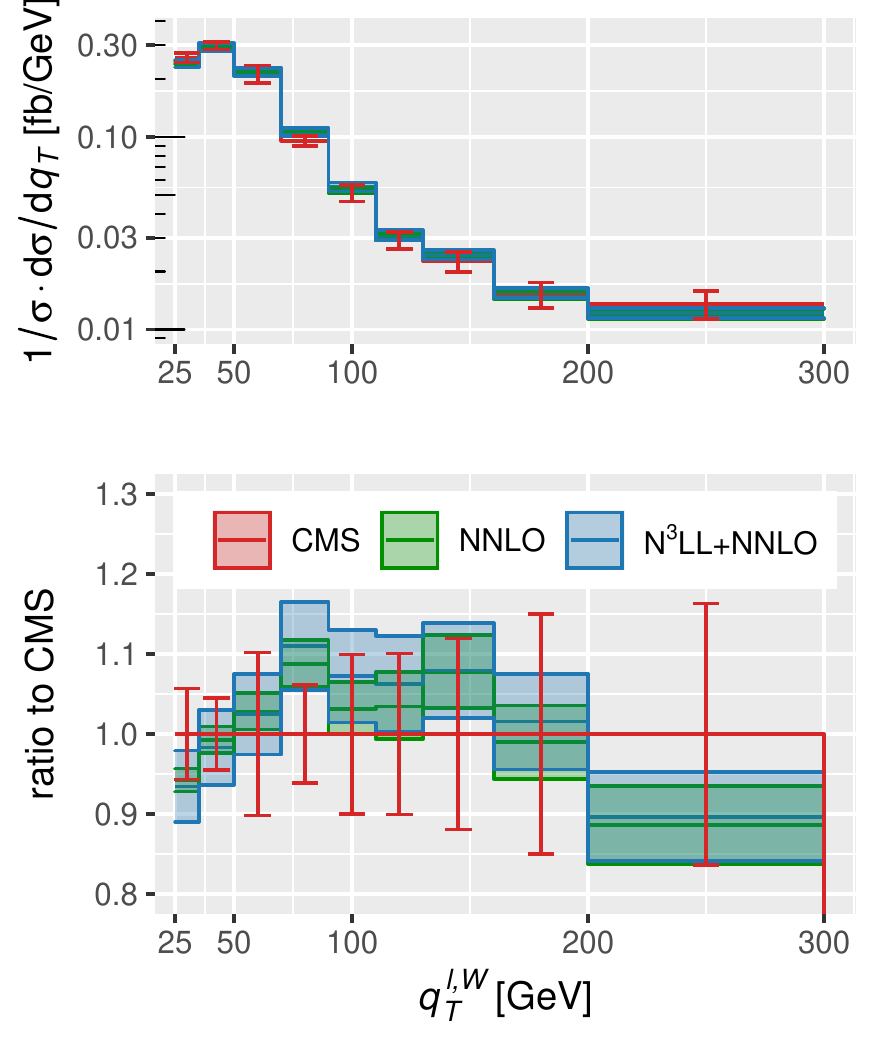}
\caption{The $q_T^{\ell_W}$ distribution for $W^\pm Z$ at \NNLO{} and \NTHREENNLO{} compared to the 
\CMS{} data from ref.~\cite{CMS:2021icx}.}
\label{fig:WZCMSptl}
\end{figure}

We now turn to the differential comparison with the measurement of ref.~\cite{CMS:2021icx}.
The transverse momentum distribution of the lepton in the $W$ decay (summed over
both $W$ charges) $q_T^{\ell_W}$ is shown at \NNLO{} and \NTHREENNLO{} in \cref{fig:WZCMSptl}.
Resummation appears to have no significant effect for this variable.

\subsection{$W^+W^-$ production} \label{sec:WWproduction}

The experimental study of $W^+W^-$ production is subject to large backgrounds,
principally from top production, but also from Drell-Yan processes, $W$+jet production,
and other di- and tri-boson production processes. Reducing the background from top production  
to an acceptable level currently requires the imposition of a veto on jet activity. We have 
implemented jet-veto resummation for all single boson and boson pair processes
at the level of \NTHREENNLOp{} based on the collinear anomaly formalism 
\cite{Becher:2013xia} and the ingredients available in the literature 
\cite{Abreu:2022sdc,Abreu:2022zgo}. Here we only present results for 
$W^+ W^-$ production and leave a detailed jet-veto study for a future publication \cite{CENS}. 
Previous detailed analyses of this process in the literature are at the level of \NNLL{} 
\cite{Jaiswal:2014yba,Arpino:2019fmo,Kallweit:2020gva} and \NTHREENNLOp{}
\cite{Dawson:2016ysj}. Our implementation of the jet veto resummation is closer to 
full \NTHREENNLO{} than ref.~\cite{Dawson:2016ysj}, since it contains complete results for the beam function~\cite{Abreu:2022zgo},
including dependence on the jet radius~$R$.

\begin{table}
	\caption{\label{tab:WW-CMS} Setup for the 
		\CMS{} $W^+W^-$ fiducial volume analysis at $\sqrt{s}=\SI{13}{\TeV}$ presented in 
		ref.~\cite{CMS:2020mxy}.}
	\vspace{0.5em}
	\begin{center}
		\begin{tabular}{r | l }
			{ lepton cuts}
			& $q_T^{\ell} > \SI{20}{\GeV}$, $|\eta^{\ell}|<2.5$, \\
			& $m_{\ell\ell}> \SI{20}{\GeV}$, $q_T^{\ell\ell}>\SI{30}{\GeV}$, \\
			&  $q_T^{\text{miss}}> \SI{30}{\GeV}$\\
			{ jet veto} & anti-$k_T$, $R=0.4$, 0-jet events only
		\end{tabular}
	\end{center}
\end{table}

\begin{figure} 
	\centering
	\includegraphics[width=0.85\columnwidth]{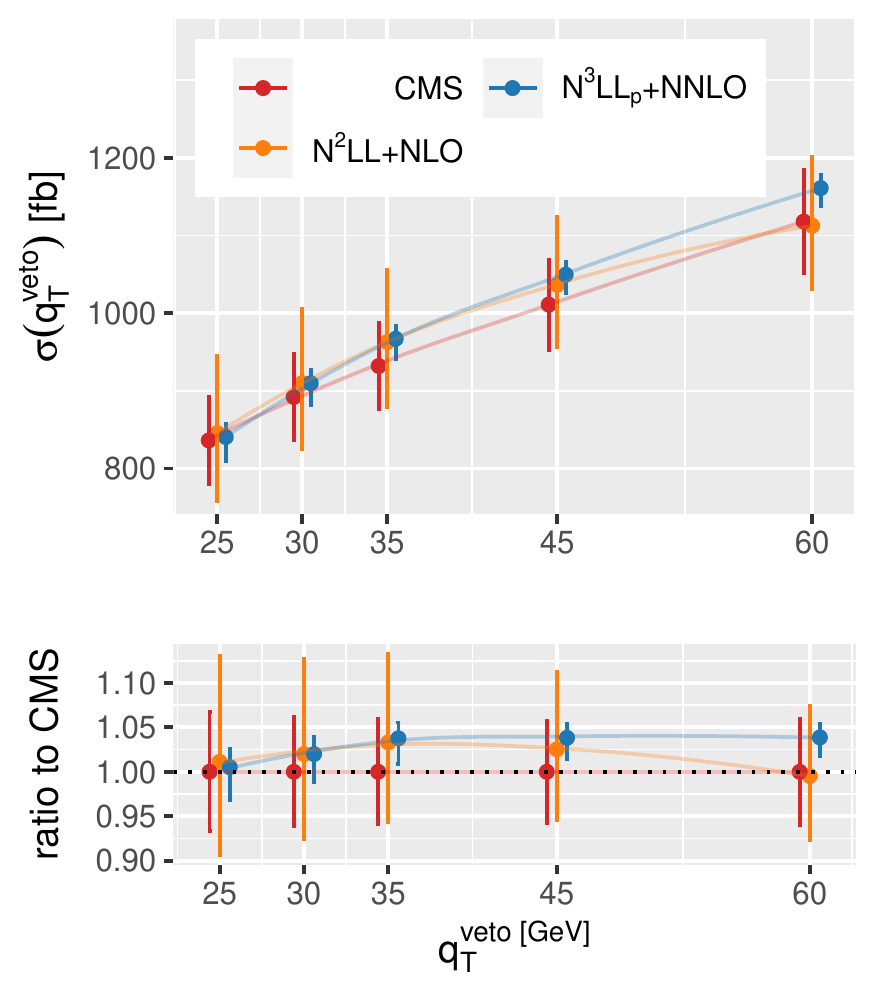}
	\caption{Jet-veto resummed cross-sections for $W^+W^-\to 2e2\nu$ production at \SI{13}{\TeV} 
		using the \CMS{} cuts in \cref{tab:WW-CMS} of this section in comparison with the \CMS{} 
		measurement \cite{CMS:2020mxy}. The solid lines are interpolations to 
		guide 
		the eye.}
	\label{fig:WW13qt}
\end{figure}

In \cref{fig:WW13qt} we present jet-veto resummed results at $\sqrt{s}=\SI{13}{\TeV}$ using the 
\CMS{} cuts in \cref{tab:WW-CMS} compared with the corresponding analysis 
\cite{CMS:2020mxy}. The calculation is performed using the 
$n_f=4$ version of the \PDF{}s to ensure consistency across the entire calculation. We include a
factor of 4 in our results to account for the sum over both electrons and muons, neglecting 
contributions from $ZZ \to \ell \bar\ell \nu\bar\nu$ that are negligible as long as a suitable cut 
on $|m_{\ell\bar\ell}-m_Z|$ is applied. We include the $gg$ channel at leading fixed 
order. For simplicity we do not include a transition function for this calculation but perform a 
naive matching.
The cross-section is dominated by the resummed part, with 5 to 10 percent matching
corrections between $q_T^\text{veto}=\SI{25}{\GeV}$ and \SI{60}{\GeV}. Overall there is agreement 
between our theory predictions and the measurements within uncertainties.

\subsection{$WH$ and $ZH$ production} \label{sec:VHproduction}
Matched \NTHREENNLO{} calculations for $WH$ and $ZH$ were implemented in ref.~\cite{Becher:2020ugp},
but no results were presented. Here we present predictions for these two processes
at $\sqrt{s}=\SI{13.6}{\TeV}$ to demonstrate the capabilities of the code. For this demonstration 
we do not 
apply cuts on the electroweak final state after $W$, $Z$ and $H$ bosons decay. We further divide 
out the branching ratio to give the total rate independent of the particular decay channel.

We have also upgraded our code to include resummation for $W\gamma$ production, but do not show 
results, since $Z\gamma$ has been extensively discussed in ref.~\cite{Becher:2020ugp}. In 
particular, the issue of photon isolation plays a big role in this process and requires a dedicated 
discussion.

\begin{figure}
	\centering
\includegraphics[width=0.85\columnwidth]{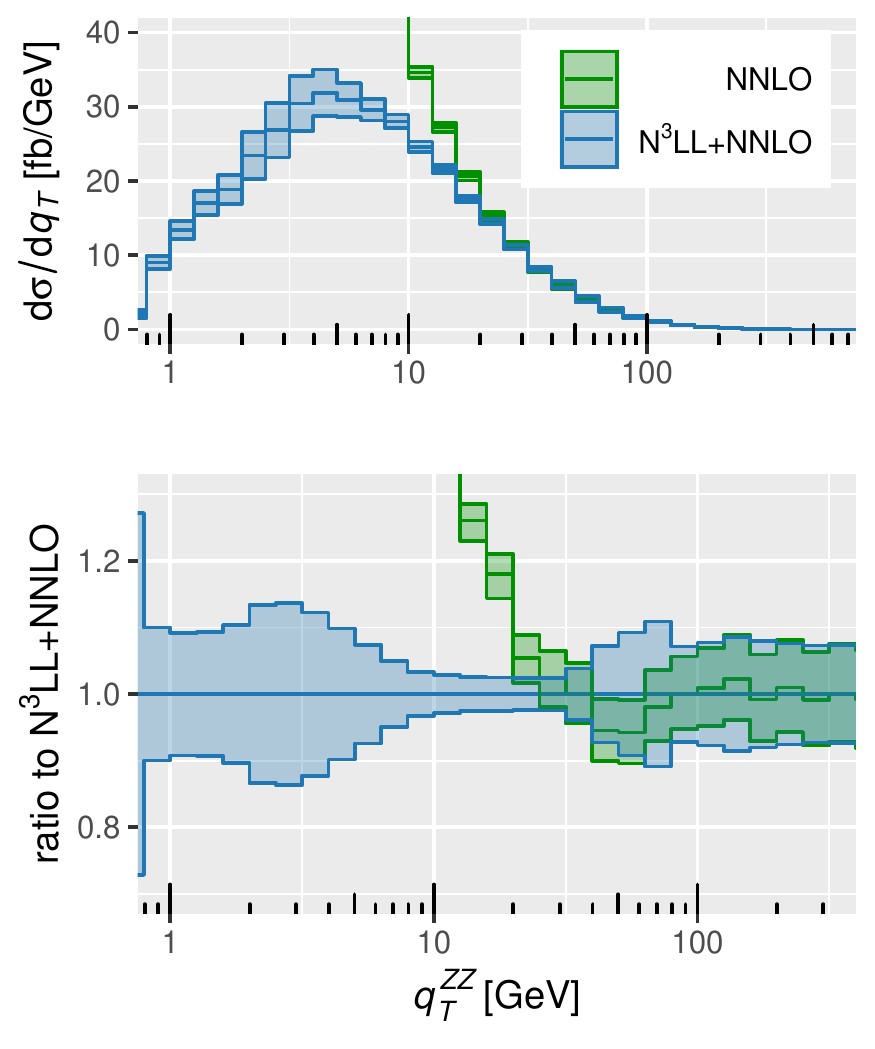}
\caption{The $q_T^{WH}$ distribution for $W^+H+W^-H$ at \NTHREENNLO{} compared to fixed order 
\NNLO{}.}
\label{fig:WH136qt}
\end{figure}

\begin{figure}
	\centering
\includegraphics[width=0.85\columnwidth]{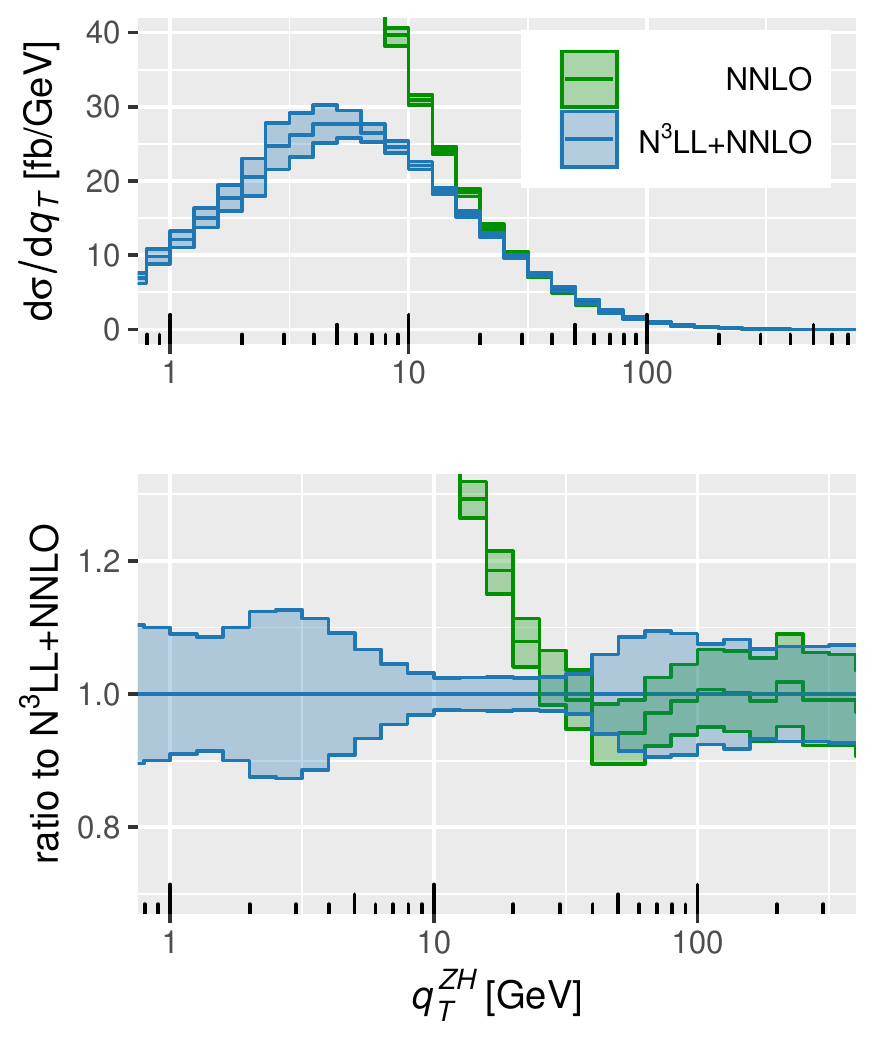}
\caption{The $q_T^{ZH}$ distribution at \NTHREENNLO{} compared to fixed order \NNLO{}.}
\label{fig:ZH136qt}
\end{figure}

\section{Conclusions}

The experimental study of massive diboson kinematics is still in its
infancy. Currently only $5\%$ of the final \LHC{} luminosity has been recorded and the
high-luminosity \LHC{} will require precise predictions at the percent level.
We have presented transverse momentum resummed results at the level of \NTHREENNLO{} for the 
production of pairs of vector bosons $ZZ$, $W^{\pm}Z$, $W^{\pm}H$ and $ZH$. 
Where experimental data has been available in sufficient detail we have shown that the
inclusion of the resummed logarithms leads to improved agreement with the data at low $q_T$. For 
$W^+W^-$ production we have shown jet-veto resummed predictions in comparison with measurements and 
find agreement within uncertainties.

Current binning of experimental data is not fine-grained and still quite inclusive, in particular one often has a large first 
bin starting at $q_T=0$. This diminishes the effect of $q_T$ resummation, which is most necessary 
at the differential level at small $q_T$, but also for certain sets of fiducial cuts at an 
inclusive level at a sufficient level of precision \cite{Salam:2021tbm}.
Our more finely binned predictions for $\sqrt{s}=\SI{13.6}{\TeV}$ show the importance of  
resummation when precise enough data becomes available.

With decreasing experimental uncertainties it will be necessary to take into account \NLO{} 
corrections to the $gg$ channel \cite{Grazzini:2020stb,Grazzini:2018owa}, which we only include at 
\LO{}, as well as 
\NLO{} electroweak corrections \cite{Kallweit:2014xda,Kallweit:2017khh,Grazzini:2019jkl} that can
be included by the use of automated one-loop 
tools~\cite{Actis:2016mpe,Chiesa:2015mya,Kallweit:2014xda}.
Further refinements are possible
through the inclusion of identical-particle effects, which we have neglected so far but are
straightforward to include once experimental results become sufficiently precise.

Our calculation will be publicly available in the upcoming release of \MCFM{} and can be used to 
reproduce the results in this study as well as to perform further studies with modified parameters. 
With this, our calculation also provides 
an important theoretical tool for comparison and tuning of approaches based on parton shower event 
generators operating at low logarithmic accuracy.

\paragraph{Acknowledgements.}

This manuscript has been authored by Fermi Research Alliance, LLC under Contract No. DE-AC02-07CH11359
with the U.S. Department of Energy, Office of Science, Office of High Energy Physics (JMC).
TN is supported by the United States Department of Energy under Grant Contract DE-SC0012704. This research used 
resources of the National Energy Research Scientific
Computing Center (NERSC), a U.S. Department of Energy Office of Science
User Facility located at Lawrence Berkeley National Laboratory, operated
under Contract No. DE-AC02-05CH11231 using NERSC award HEP-ERCAP0021890.

\appendix

\bibliography{main}
\bibliographystyle{JHEP}

\end{document}